\documentclass[useAMS,usenatbib]{mnras}

\usepackage{amsmath}
\usepackage{amssymb}
\usepackage{graphicx}
\usepackage{dcolumn}
\usepackage{bm}
\usepackage{ulem}

\title[Breaking properties of neutron star crust]
{Breaking properties of neutron star crust}

\author[D.A. Baiko and A.I. Chugunov]
{D.A. Baiko\thanks{E-mail:baiko@astro.ioffe.ru} 
and A.I. Chugunov\thanks{E-mail:andr@astro.ioffe.ru} \\
Ioffe Institute, Politekhnicheskaya 26, 194021 Saint Petersburg, 
Russian Federation}

\begin{document}

\label{firstpage}
\date{Accepted; Received ; in original form}

\pagerange{\pageref{firstpage}--\pageref{lastpage}} \pubyear{2018}

\maketitle

\begin{abstract}
%
The strength of neutron star crust is crucial for modelling magnetar 
flares, pulsar glitches and gravitational wave emission. We 
aim to shed some light on this problem by
analysing uniaxial stretch deformation (elongation and contraction) of 
perfect body-centered cubic Coulomb crystals, paying special 
attention to the inherent anisotropy of this process. Our analysis 
is based on the semi-analytical approach of Baiko \& Kozhberov (2017), 
which, for any uniform deformation, allows one to calculate, in fully 
non-linear regime, critical deformation parameters beyond which the 
lattice loses its dynamic stability. We 
determine critical strain, pressure anisotropy and 
deformation energy for any stretch direction with respect to the 
crystallographic axes. These quantities are 
shown to be strongly anisotropic: they vary by a factor of almost 10 
depending on the orientation of the deformation axis. For 
polycrystalline crust, we argue that the 
maximum strain for the stretch deformation sustainable elastically 
is 0.04. It is lower than the breaking strain of 0.1 obtained in 
molecular dynamic simulations of a shear deformation by 
Horowitz \& Kadau (2009). The maximum pressure anisotropy of 
polycrystalline matter is estimated to be in the 
range from 0.005 to 0.014 $nZ^2e^2/a$, where $n$ is the ion number 
density, $Ze$ is the ion charge, and $a$ is the ion-sphere radius. 
We discuss possible mechanisms of plastic motion and 
formation of large crystallites in neutron star crust as well as 
analyse energy release associated with breaking  
of such crystallites in the context of magnetic field evolution 
and magnetar flaring activity.
%
\end{abstract}


\section{Introduction}
The crust of a neutron star (NS) is crystallised, i.e.\ ions (atomic 
nuclei) form a lattice (e.g., \citealt*{hpy07,ch08,ch17}). In most 
of the crust the groundstate lattice has the body-centered cubic (bcc) 
structure (see \citealt*{Baiko14} for discussion of screening effects).
Crystalline properties of NS crust are invoked to explain a number of 
observed astrophysical phenomena. For instance, crustal 
elasticity is thought to be important for interpretation of 
quasi-periodic oscillations observed in magnetars (e.g., 
\citealt*{Gabler_etal18}). Solid crust can support 
asymmetric distributions of density, so-called, mountains, which can 
produce gravitational wave (GW) emission 
(\citealt*{ucb00,Horowitz10,jmo13}). 
Size of these mountains is limited by the strength of the 
crust (i.e.\ maximum stress, which the elastically deformed crustal 
material can withstand). If the crust is indeed as strong as it 
is suggested by molecular dynamic (MD) simulations 
(\citealt*{hk09,ch10,ch12}), the GW emission can be poweful enough to 
be detectable by existing GW observatories (\citealt*{jm13,ch17}). 
\cite{hp17} have recently suggested 
that this GW emission could be responsible for observed spin-down 
during the accreting phase of PSR J1023+0038. Crust breaking under 
magnetic stress is likely responsible for magnetar outbursts 
(e.g., \citealt*{llb16}), while plastic motion may be crucial for 
magnetic field evolution (e.g., \citealt*{Lander16}). 
\cite{Tsang_etal12} argued that crust breaking could produce an 
electromagnetic precursor of the GW signal from NS mergers. 

Thus, an accurate knowledge of the crust strength is  
fundamental for NS physics and in this paper we analyse it, paying 
special 
attention to its anisotropy. Previous works in this field explored a 
shear deformation by MD simulation (\citealt*{hk09,ch10,ch12,hh12}) and
volume-conserving crystal stretching in two highly symmetric 
directions semi-analytically (\citealt*{BK17}; Paper 1). Here we apply 
the approach proposed in the latter work to study stretching in 
arbitrary directions.

In Paper 1, the existence of a limit of a uniform 
Coulomb crystal deformation, above which the crystal loses dynamic 
stability, has been demonstrated  using standard lattice dynamics. 
The maximum crystal elongation and 
contraction factors as well as the maximum pressure anisotropy at 
breaking has been predicted essentially analytically. The authors have 
discussed two 
particular stretch directions and have found that the limiting stretch 
factor and the pressure anisotropy were strongly dependent on the angle 
between the stretch direction and the crystallographic axes. 

In this paper, we generalise the results of Paper 1 and perform an 
extensive study of the dependence of the maximum deformation, pressure 
anisotropy and deformation energy on the stretch direction for a 
perfect bcc Coulomb crystal (Section \ref{Sec_res}). We find this 
dependence to be very strong (for instance, critical pressure anisotropy 
differs by a factor of 
almost 10 between the weakest and the strongest 
directions). We point out that the maximum deformation is typically 
smaller for a contraction of the lattice, than for an elongation. 
Also, typical maximum contraction deformation is about 5\%, 
which is a factor of two lower than the breaking strain for the shear 
deformation, obtained by \cite{hk09} and \cite{ch10,ch12}. Our results 
clearly demonstrate that the Tresca and von Misses failure criteria 
are not valid for perfect Coulomb lattice in NS crust. Our limits of 
the lattice stability are upper bounds for particular deformation types, 
because other factors, such as ion motion about lattice nodes (due to 
zero-point or finite temperature effects) or electron screening, will 
further degrade material strength. 

In Subsection \ref{Sec_elastEnerg} we calculate the elastic energy at 
the critical deformation for all stretch directions.

In Section \ref{Sec_spec} we consider astrophysical applications 
of our results. In particular, Subsection \ref{Sec_poly} is devoted to 
breaking properties of polycrystalline matter, which is a possible 
microscopic state of NS crust.
In Subsection \ref{Sec_Heat} we estimate 
heat sources associated with breaking events in magnetar crust. 

Our results can be applied to both outer and inner crust of neutron 
stars provided that the effect of dripped neutrons on the interionic 
interaction can be neglected, and excluding the exotic 
pasta phases, which may appear at mass densities exceeding 
$10^{14}$ g cm$^{-3}$.

\section{Crustal model, scalings and breaking properties}
\label{Sec_res}
We model NS crust as a deformed bcc Coulomb crystal composed of 
identical ions. This means that ions interact with each other via
pure Coulomb forces and we neglect screening of this interaction
by electrons. It is a widely employed approximation 
in NS literature (e.g., \citealt{hpy07,ch08}). In particular, it 
is known to predict elastic properties of the crust with reasonable 
accuracy (\citealt*{Baiko11,Baiko15}). \cite{ch12} have analysed effects 
of electron screening on the breaking stress for the shear deformation 
and have demonstrated that screening reduced strength, but not very 
much. Thus, we expect that an accurate treatment of electron screening 
can only reduce maximum pressure anisotropy. We discuss stability of a 
deformed {\it static} lattice and do not include any effects associated 
with ion motion, which are expected to further weaken critical 
deformation properties of the lattice (e.g., \citealt*{rst72}).

We are interested in the maximum deformations for arbitrary stretch 
directions. 
However, the symmetry of the lattice allows one to restrict themselves 
by deformations in a narrower solid angle which contains all 
essentially different  
directions of the crystal stretch. Maximum deformations for all other 
directions can then be obtained by simple symmetry transformations.
For the bcc lattice, it means studying stretch directions characterised 
by unit vectors $\bm{s}=(s_x,s_y,s_z)$ satisfying 
$s_x \geq s_y \geq s_z \geq 0$, where $xyz$ is a Cartesian reference 
frame aligned with the bcc lattice cube. We 
specify these directions by two angles, $\theta$ and $\phi$, where 
$\theta$ is the angle between the stretch direction and the $x$-axis, 
while $\phi$ is the azimuthal angle in the plane perpendicular to the 
$x$-axis. Angle $\phi$ varies from 0 [e.g., for direction 
$(1,1,0)/\sqrt{2}$] to $\pi/4$ [e.g., for direction 
$(1,1,1)/\sqrt{3}$]. Angle $\theta$ varies from 0 to
$\tan^{-1}{(1/\cos{\phi})}$.  

The stretches under study are volume-conserving deformations of the 
perfect lattice, in which all projections of the lattice vectors 
along the stretch axis are multiplied by a factor $1+\epsilon$, while 
all projections in the plane perpendicular to this direction are 
multipled\footnote{For small deformations, the Cauchy's 
strain tensor $u_{ij}$ is diagonal in the reference frame, where axis 
3 and the stretch axis coincide, 
$u_{11}=u_{22}=-\epsilon/2;\ u_{33}=\epsilon$.} 
by $1/\sqrt{1+\epsilon}$. Alternatively (as in Paper 1), 
all projections of the lattice vectors along the stretch axis can be 
multiplied by a factor $\xi$ and then an overall scale factor 
$\xi^{-1/3}$ is apllied to preserve the volume. Thus 
$\xi=(1+\epsilon)^{3/2}$. Clearly, an elongation 
corresponds to $\epsilon>0$ or $\xi>1$, while a contraction 
is obtained in the opposite case. The maximum deformation is determined 
for each direction $(\theta,\phi)$ independently for both elongation 
and contraction. It corresponds to a minimum value of $|\epsilon|$, at 
which the lattice is no longer stable or, to be more precise, at which 
a zero frequency phonon mode appears. To locate this transition, we 
calculate the dynamic matrix of the deformed Coulomb lattice using 
the Ewald transformation and solve the respective secular equation
on a dense grid of wavevectors $\bm{k}$ in the first Brillouin zone. 
For all considered deformations, the unstable mode appears 
at $k \rightarrow 0$, indicating that the loss of stability takes 
place on large spatial scales.

For certain non-volume-conserving deformations we can adopt 
a two-step procedure in which the first step is a uniform 
compression/expansion to the desired density followed by 
a volume-conserving stretch to the final state. 
Since the first  deformation preserves bcc lattice, it is always 
stable. Thus the total deformation will be stable if and only if the 
second deformation does not lead to the loss of stability.

In the Coulomb crystal, the electrostatic part of the stress 
tensor $\tilde\sigma_{ij}$ has a simple scaling with density and 
composition:
\begin{equation}
\tilde\sigma_{ij}=\frac{nZ^2 e^2}{a}\,\sigma_{ij}~, 
\label{sig_scale}
\end{equation}
where the dimensionless stress tensor $\sigma_{ij}$ depends only on 
deformation, i.e.\ on $\theta$, $\phi$, and $\xi$. 
In this case, $n$ is the ion number density, $Z$ is the ion charge 
number, $e$ is the elementary charge, and $a=(4 \pi n/3)^{-1/3}$ is 
the ion-sphere radius. For non-deformed bcc lattice, 
$\sigma_{ij} = -\zeta \delta_{ij}/3$, where $\zeta \approx -0.8959293$
is the Madelung constant (e.g., \citealt{hpy07}). 

For deformed lattice, the stress tensor remains symmetric 
($\sigma_{ij}=\sigma_{ji}$, see, e.g., \citealt*{ll_elast}) and thus
can be always cast in the diagonal form. However, it
becomes anisotropic (i.e.\ $\sigma_{ij} \ne \sigma \delta_{ji}$)
and we, in analogy with the Tresca criterion, characterise it by the 
difference between the maximum and the 
minimum eigenvalues of the stress tensor, which we refer to as pressure 
anisotropy 
$\Delta \tilde P$. According to (\ref{sig_scale}), 
\begin{equation}
 \Delta \tilde P=\frac{nZ^2 e^2}{a}\,\Delta P~, 
\label{dP_scale}
\end{equation}
where $\Delta P$ is the dimensionless pressure anisotropy. 

The excess energy of the deformed lattice with respect to the 
non-deformed configuration can be written as
\begin{equation}
	 \Delta \tilde U=\frac{N Z^2 e^2}{a}\,\Delta U~,  
\label{U_scale}
\end{equation}
where $N$ is the number of ions. 
Dimensionless quantities $\Delta P$ and $\Delta U$ 
depend neither on density nor on composition being functions of
deformation only. Note also, that $\Delta U$ is the excess energy 
per ion in units of $Z^2 e^2/a$. 

Below we present the values of $\Delta P_{\rm crit}$ and 
$\Delta U_{\rm crit}$ to 
characterise lattice properties at the onset of lattice instability. 
Scalings (\ref{dP_scale}) and (\ref{U_scale}) allow one to recalculate 
the respective properties in physical units for arbitrary density 
and composition.

\begin{figure}                                           
	\begin{center}                                              
		\leavevmode                                                 
		\includegraphics[height=73mm,bb=3 12 363 347,clip]{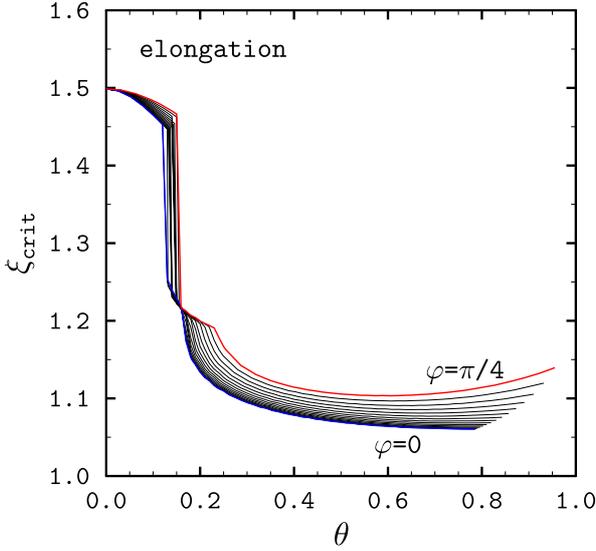} 
	\end{center}                                                
	\vspace{-0.4cm} 
	\caption[]{Critical stretch factor for elongation.}                                             
	\label{xicrit_s}
\end{figure}
%

\begin{figure}                                           
	\begin{center}                                              
		\leavevmode                                                 
		\includegraphics[height=73mm,bb=1 12 363 347,clip]{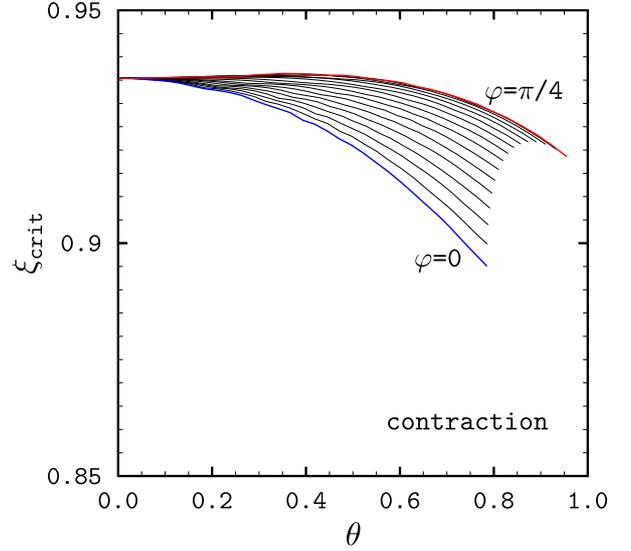} 
	\end{center}                                                
	\vspace{-0.4cm} 
	\caption[]{Critical stretch factor for contraction.}                                             
	\label{xicrit_c}
\end{figure}
%

\begin{figure}                                           
	\begin{center}                                              
		\leavevmode                                                 
		\includegraphics[height=73mm,bb=4 12 363 344,clip]{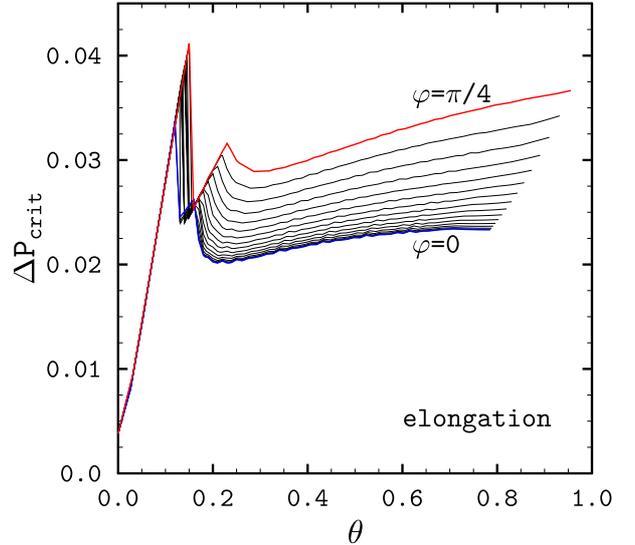} 
	\end{center}                                                
	\vspace{-0.4cm} \caption[]{Dimensionless critical pressure anisotropy 
for elongation.}                                             
	\label{dP_s}
\end{figure}
%

\begin{figure}                                           
	\begin{center}                                              
		\leavevmode                                                 
		\includegraphics[height=73mm,bb=4 12 363 344,clip]{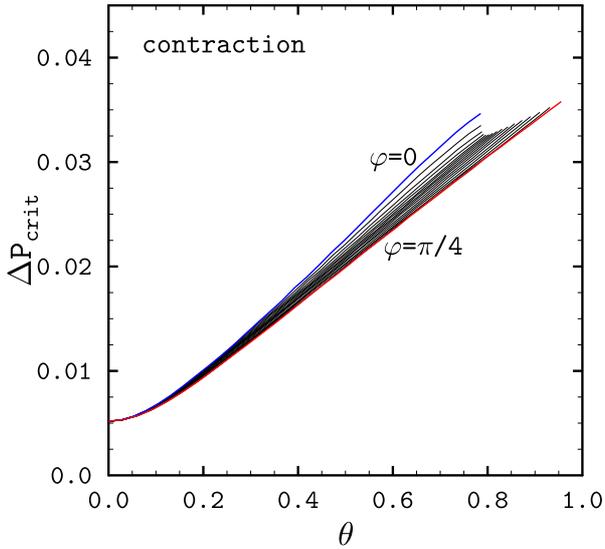} 
	\end{center}                                                
	\vspace{-0.4cm} \caption[]{Dimensionless critical pressure anisotropy 
for contraction.}                                             
	\label{dP_c}
\end{figure}
%

Our results for critical stretch factors $\xi_{\rm crit}$ are 
summarised in Fig.\ \ref{xicrit_s} for elongation and 
Fig.\ \ref{xicrit_c} for 
contraction. In both cases we show the dependence of the critical 
stretch factor on $\theta$ for 16 values of $\phi$ 
equally spaced between 0 (blue curve) and $\pi/4$ (red curve). 
In Figs.\ \ref{dP_s} and \ref{dP_c} we show critical pressure 
anisotropy $\Delta P_{\rm crit}$ for elongation and contraction, 
respectively, as a function of the same angles. This quantitity is 
calculated using the Ewald transformation for the critically deformed
Coulomb lattice. 

Let us start with a discussion of elongation. One sees that the 
critical strain strongly depends on $\theta$: for $\theta \lesssim 0.2$, 
it can be as large as $\epsilon_\mathrm{crit} \approx 0.3$ 
($\xi_{\rm crit} \approx 1.5$), but for $\theta \gtrsim 0.4$, 
$\epsilon_\mathrm{crit} \lesssim 0.07$ ($\xi_{\rm crit} \lesssim 1.1$). 
This is 1.4 times smaller than the critical strain 
$\epsilon^{\rm MD}_{\rm crit} \approx 0.1$ obtained in MD simulations 
of the shear deformation (\citealt*{hk09}). 
Weakness (i.e.\ loss of stability at small $\Delta P$) of the Coulomb 
lattice for stretching along the cube edge 
($\theta=0$) was demonstrated in Paper 1; our results reveal that 
it takes place in a rather narrow solid angle. The critical 
pressure anisotropy $\Delta P_{\rm crit}$ rapidly increases with 
increase of 
$\theta$. For $\theta \gtrsim 0.1$, $\Delta P_{\rm crit}$ varies 
within a factor of 2 ($\Delta P_{\rm crit} =$ 0.02--0.04). 
In the vicinity of the global maximum, it reaches the 
breaking stress for the shear deformation of 
a perfect crystal, $\Delta P^{\rm MD} \approx 0.039$, obtained by 
\cite{ch10} at $T \rightarrow 0$. 
Dependence of $\epsilon_\mathrm{crit}$ and $\Delta P_{\rm crit}$ 
on $\phi$ is not 
strong (within a factor of 2 between the minimum and the maximum values 
for a given $\theta$).

Critical parameters for contraction exhibit vastly different behaviour. 
The critical deformation increases monotonically with 
increase of $\theta$. Dependence on $\phi$ is also monotonic. 
For instance, $\epsilon_\mathrm{crit} \approx -0.04$ 
($\xi_{\rm crit} \approx 0.935$) at $\theta=0$ 
(and arbitrary $\phi$), but at $\theta\sim 0.8$, one has 
$\epsilon_\mathrm{crit} \approx -0.07$ 
($\xi_{\rm crit} \approx 0.895$) for $\phi=0$  
and $\epsilon_\mathrm{crit} \approx -0.05$ 
($\xi_{\rm crit} \approx 0.92$) for $\phi=\pi/4$. It is 
worth emphasising, that, for contractions, minimum 
$|\epsilon_\mathrm{crit}|$ 
obtained here is 2.5 times below\footnote{There is no direct 
contradiction between these data, 
because, being presented in eigen-axes $1^\prime 2^\prime 3^\prime$, 
the shear deformation ($u_{12}=u_{21}=\epsilon/2,\ u_{33}=0$) 
corresponds to $u_{1^\prime 1^\prime }=\epsilon/2,\ 
u_{2^\prime 2^\prime }=-\epsilon/2,\ u_{3^\prime 3^\prime }=0$. That is,
it corresponds to a deformation anisotropic in the plane, perpendicular 
to the stretch axis. Such deformations are beyond the scope of 
the present paper. \label{foot_shear}}  
$\epsilon^{\rm MD}_{\rm crit} \approx 0.1$.
The critical pressure anisotropy increases with increase of $\theta$ 
(from $\Delta P_{\rm crit} \approx 0.005$ at $\theta=0$ to 
$\Delta P_{\rm crit} \approx 0.035$ at $\theta\sim 1$). The dependence 
on $\phi$ is also not very strong 
(up to 15\% between the minimum and the maximum values). For all 
$\theta$ and $\phi$, the crystal breaks at lower pressure anisotropy 
than the critical $\Delta P^{\rm MD} \approx 0.039$. 

For $\theta=0$, 
$\Delta P_{\rm crit} \sim 0.005$ for both elongation and contraction, 
but for other stretching directions, the lattice is generally asymmetric 
for elongation/contraction.

\subsection{Elastic energy}
\label{Sec_elastEnerg}
\begin{figure}                                           
\begin{center}                                              
\leavevmode                                                 
\includegraphics[height=73mm,bb=12 12 363 347,clip]{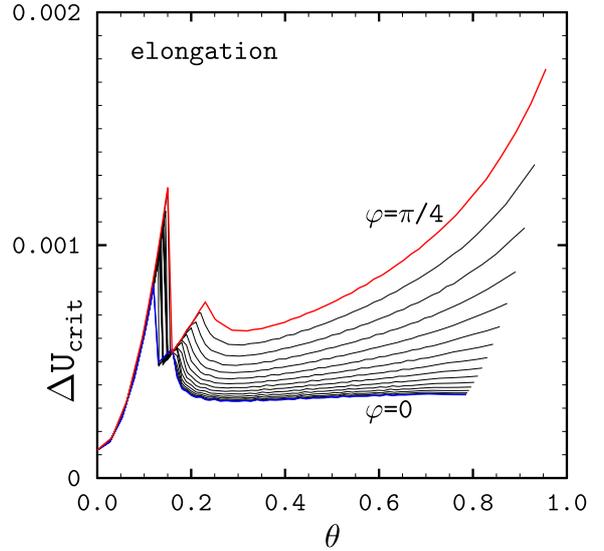} 
\end{center}                                                
\vspace{-0.4cm} \caption[]{Dimensionless excess electrostatic energy 
per ion at critical elongation.}
\label{dU_s}
\end{figure}

\begin{figure}                                           
\begin{center}                                              
\leavevmode                                                 
\includegraphics[height=73mm,bb=12 12 363 344,clip]{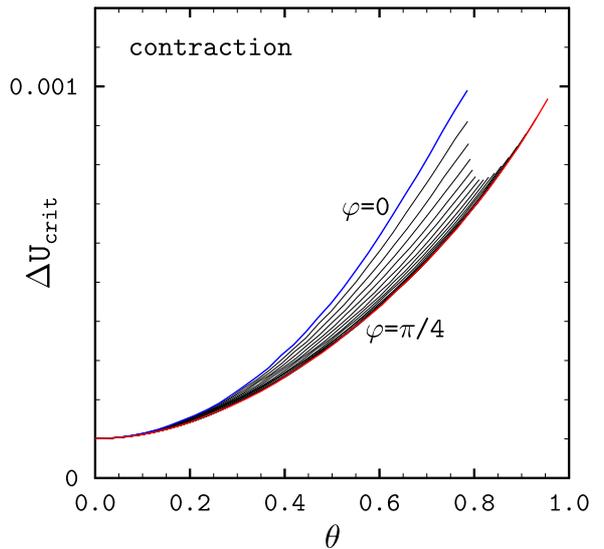} 
\end{center}                                                
\vspace{-0.4cm} \caption[]{Dimensionless excess electrostatic energy 
per ion at critical contraction.}
\label{dU_c}
\end{figure}

Stretching the crystal (e.g., by evolving magnetic field) requires 
work against the elastic forces and increases the crystal electrostatic 
energy. The latter can be easily computed using the same Ewald 
method as that employed for the stress 
tensor calculation\footnote{Non-linear behaviour of the stress-strain 
curve before the loss of stability (see Fig.\ 1 in Paper 1) does not 
allow one to apply the simple linearised theory to calculate the elastic 
energy precisely.}. The excess energies at critical deformations are 
shown in Figs.\ \ref{dU_s} and \ref{dU_c} for elongation and 
contraction, respectively, as functions of the same angles. Similar to 
$\Delta P_{\rm crit}$, $\Delta U_{\rm crit}$ is strongly anisotropic 
and demonstrates 
a monotonic behaviour as a function of $\theta$ only for contraction. 
For elongation, the minimum of $\Delta  U_{\rm crit}$ at $\theta=0$ is 
less pronounced than the minimum of $\Delta P_{\rm crit}$. This is due 
to strongly non-linear elasticity at $\theta \approx 0$ 
(see Fig.\ 1 of Paper 1). Let us point out, that 
for all considered deformations, $\Delta U_{\rm crit}$ is insufficient 
to completely melt 
the crystal, even for a crystal that is already at melting 
temperature, $T_{\rm m}$, because the latent heat for a Coulomb crystal 
is expected to be greater ($\sim 0.75 kT_{\rm m} \sim 0.004 Z^2 e^2/a$ 
neglecting quantum effects, e.g., \citealt*{H04}).  

%

%

\section{Applications} 
\label{Sec_spec}   

\subsection{Polycrystalline matter} 
\label{Sec_Def_Poly}
Let us adopt a popular point of view that matter in the 
NS crust is polycrystalline (e.g., \citealt*{Strohmayer_etal91,ch08}). 
Although the 
validity of this picture is not guaranteed, we defer a discussion of an 
alternative point of view till Subsection \ref{Sec_Heat}. 
\cite{Caplan_etal18} discuss crystallisation of matter in accreting NS  
and argue that the typical size of crystallites at the crust 
top should be comparable with the scale height $\sim 10$~m. 
Subsequent compression by pressure of newly accreted material 
should decrease crystallite (grain) size (if there is no merging 
between them --- see Subsection \ref{Sec_Plastic}), producing a 
polycrystal with even smaller grains. 
 
If crystallites are randomly oriented and are 
much smaller than the typical scale of stress and strain variation, 
the infinitesimal deformations of the polycrystal can be described by 
the isotropic linear theory of elasticity, in which a material 
is characterised by just two quantities, the bulk modulus $K$ and 
the effective shear modulus $\mu_\mathrm{eff}$. However, these 
quantities depend on the correlations between the orientations of the 
neighbouring crystallites, and, in general, cannot be accurately 
calculated from elastic properties of a monocrystal 
(e.g., \citealt{ll_elast}). 

Several ways to determine $\mu_\mathrm{eff}$ have been proposed 
(see, e.g., \citealt*{kp15} for review) with numerical results 
diverging by as much as a factor of 2.34 assuming bcc crystallites 
and neglecting electron screening. In NS literature, one typically 
uses the effective shear modulus 
$\mu_\mathrm{eff}^\mathrm{V} \approx 0.1194$ (times $n Z^2 e^2 / a$ 
in physical units) introduced by \cite{oi90} on the basis of an angular 
averaging of the shear-wave velocity. The same quantity is obtained
if one assumes a uniform strain in all crystallites and is known as the 
Voigt average. If, on the contrary, a uniform stress in all  
crystallites is assumed, one derives the Reuss average 
$\mu_\mathrm{eff}^\mathrm{R}\approx 0.051$. \cite{kp15} have 
proposed another value, based on the self-consistent theory 
of \cite{Eshelby61}\footnote{For strain $|\epsilon| \lesssim 0.04$ the 
stress-strain curve for the polycrystalline MD simulation in Fig.\ 1 
of \cite{hk09} agrees well with the results of \cite{oi90}, but not 
with those of \cite{kp15}. However, this can be a specific property 
of this particular polycrystal. Note, that for larger $|\epsilon|$, 
the stress-strain curve becomes strongly non-linear, which can be an 
evidence of yield.}.

\subsection{Breaking properties of polycrystalline matter} 
\label{Sec_poly}
The maximum deformation, which can be elastically supported by the 
polycrystalline crust, is even more uncertain. 
Previously, various authors (e.g., \citealt{ucb00,jmo13}) 
applied the von Misses criterion by introducing
$\sigma^2 \equiv \hat\sigma_{ij}\hat\sigma^{ji}/2$,
where $\hat \sigma_{ij}=\sigma_{ij}-\sigma_{ll}\delta_{ij}/3$ was 
the traceless stress tensor, and assumed that the critical
value of $\sigma$ was  
$\sigma^{\rm trad}_{\rm max} \approx \mu^{\rm V}_\mathrm{eff} 
\epsilon^{\rm trad}_\mathrm{crit}$. Following \cite{hk09}, who have 
performed the first direct MD simulations of the breaking strain for 
the shear deformation, $\epsilon^{\rm trad}_\mathrm{crit}$ was 
typically set equal to $\epsilon^{\rm MD}_{\rm crit} \approx 0.1$ 
(e.g., \citealt{llb16}). \cite{hk09} have performed simulations not 
only for monocrystals, but have also made one simulation for a 
polycrystal and several simulations for monocrystals with defects. They 
have concluded that the resulting critical strains were only modestly 
reduced in comparison with $\epsilon^{\rm MD}_{\rm crit}$.

Our analysis indicates that, for monocrystals, the breaking parameters 
are strongly anisotropic and cannot be 
described in terms of a single number. But what does this tell us 
about polycrystalline crust? Just like for infinitesimal deformations, 
an accurate answer requires modelling the stress/strain 
distribution inside crystallites and depends on their specific 
geometry and mutual orientation. However, a detailed solution like that 
is simultaneously very hard to get and not very useful in the 
astrophysical context. Instead, let us  
rely on the same ``smoothed-out'' models as in Subsection 
\ref{Sec_Def_Poly}. As discussed by \cite{Kocks70}, these models are not 
fully self-consistent, however, the first one (uniform strain, Voigt) 
results in an upper bound for the strength, whereas the second one 
(uniform stress, Reuss) gives the lower 
bound, provided that the yield or breaking is associated with a loss 
of stability inside one of the crystallites.

In the first approach, the 
strength must be determined by the minimum 
(over the stretch directions) of the absolute value of the critical 
strain, $|\epsilon_{\rm crit}| \approx 0.04$ (for elongation it 
corresponds to $\theta \approx 0.8$ and $\phi \approx 0$, 
Fig.\ \ref{xicrit_s}, for contraction it occurs at $\theta \approx 0$ 
and any $\phi$, Fig.\ \ref{xicrit_c}). For larger strain, a suitably 
oriented crystallite will lose stability and the deformation will 
become inelastic. Accordingly, we expect that the critical strain for 
elongation/contraction in NS crust cannot exceed 0.04. 
Respective estimate of the critical pressure anisotropy for the 
polycrystal is $\Delta P^{\rm V}_{\rm crit} 
\approx 3 \mu_\mathrm{eff}^\mathrm{V}\epsilon_{\rm crit} \approx 0.014$
corresponding to the von Misses parameter 
$\sigma_{\rm max}^{\rm V} 
\approx \sqrt{3} \mu^{\rm V}_{\rm eff} \epsilon_{\rm crit}\approx 0.008$
[it is easy to see, e.g., from Eq.\ (4.6) of \citealt{ll_elast} that 
in the linear regime 
$\Delta P = \sqrt{3} \sigma = 3 \mu \epsilon$ for a stretch 
of an isotropic body with the shear modulus $\mu$]. 

The von Misses parameter $\sigma_{\rm max}^{\rm V}$ just obtained is 
1.5 times lower than the traditional estimate above:  
$\sigma^{\rm trad}_{\rm max} \approx 0.0119$. At the same time, it is 
in a surprisingly ideal agreement with the maximum dimensionless stress 
$\sigma^{\rm pMD}_{12,{\rm max}} \approx 0.008$ 
achieved in the polycrystalline MD simulation of \cite{hk09} [for the 
shear deformation of the isotropic body in the linear regime 
$\Delta P = 2 \sigma = 2 \mu \epsilon$, while the component 
$\sigma_{12}$ equals $\sigma$].

In terms of the pressure anisotropy, 
$\Delta P^{\rm V}_{\rm crit} \approx 0.014$ is slightly lower than 
the critical pressure anisotropy inferred from the polycrystalline MD, 
$\Delta P^{\rm pMD}_{\rm crit} \approx 2 \sigma^{\rm pMD}_{12,{\rm max}} 
\approx 0.016$; is 1.7 times lower than the traditional estimate, 
$\Delta P^{\rm trad}_{\rm crit} \approx 2 \sigma^{\rm trad}_{\rm max} 
\approx 0.0238$; and is 2.8 times lower 
than $\Delta P^{\rm MD} \approx 0.039$ for the monocrystal. 

In the second approach, 
if $\Delta P$ exceeds $0.005$, which is the 
minimum critical pressure anisotropy 
($\Delta P_{\rm crit} \approx 0.005$ at $\theta \approx 0$ and any 
$\phi$ for both elongation and contraction, Figs.\ \ref{dP_s} and 
\ref{dP_c}), the crystallite with a suitable alignment
will break. The quantity $\epsilon^{\rm R}_{\rm crit} \approx 
0.005/(3\mu_\mathrm{eff}^\mathrm{R}) \approx 0.03$ is the respective 
estimate of the critical strain.  

Notice, that the critical strains predicted by the two approaches are 
in a good agreement with each other suggesting that 
$\epsilon_{\rm crit} \approx$ 0.03--0.04 is a rather robust bound for 
the elastic stretching deformation of the polycrystalline crust. 

Let us conclude this part by reminding that effects of electron 
screening and ion motion will further reduce the critical deformation 
parameters.

\subsection{Plastic motion} 
\label{Sec_Plastic}
Finally, let us discuss mechanisms of
inelastic deformation in polycrystalline crust. 
Firstly, the loss of stability and destruction of 
a small-scale crystallite does not necessarily constitute 
a catastrophic event because the energy release will be low (cf.\ 
Subsection \ref{Sec_Heat}). The affected ions may recrystallise and
the polycrystal may survive, but become plastically deformed.
Secondly, due to anisotropy of the 
elastic properties, deformation of neighbouring grains with the same 
strain will lead to different elastic energy (per ion) in different 
grains (see Figs.\ \ref{dU_s} and \ref{dU_c} for elastic energy at 
maximum stable deformation). On top of that, boundary conditions may 
require a density variation at the grain boundary, due to which the 
scale of the electrostatic energy as well as the 
energy of the degenerate electron gas in those grains will also be 
different. It can make growth of grains with lower total energy 
at the expense of those with higher energy 
thermodynamically favourable and lead to plastic motion. Note, that 
the stress-induced grain growth is known for terrestrial materials and 
it can be a mechanism of plastic flow (e.g., 
\citealt*{grain_grow_Science2009}). 

Other mechanisms of plastic motion (e.g., grain boundary sliding, 
rotation and merging of grains, see \citealt*{GrainRotationAndSliding})
are also not excluded.  Thus, the dominating mechanism of the plastic 
motion in the NS crust is still highly uncertain and requires 
additional modelling.

\subsection{Heat source} 
\label{Sec_Heat}
Crystallisation of NS crust is an extremely slow and uniform 
process with plenty of time to anneal out any defects.
Additionally, if a large-scale sufficiently strong magnetic field is 
present, crystallisation may proceed in a way that favours 
a particular orientation of crystal axes with respect to the magnetic 
field (\citealt*{Baiko11}). As discussed in 
Subsection \ref{Sec_Plastic}, 
merging of the crystallites may occur in the course of the plastic 
deformation.  These conditions may result in a formation of large-scale 
perfect or close to perfect crystallites. Under the action of stresses 
(e.g., due to magnetic field), the crystallites will be 
stretched non-uniformly. If typical scale of the stretch-factor 
variation becomes smaller than the crystallites' size,  
the monocrystal theory developed in Section \ref{Sec_res} will  
directly apply. 

The sizes of these crystallites in NS crust will be ultimately 
determined by stresses produced by the evolving magnetic field 
(Paper 1). When the breaking limit for a crystallite segment 
is reached, a possible outcome is a destruction of this 
segment followed by a formation of a new stress-free configuration.
This is a mechanism of crystallite size reduction.
On the other hand, such a process would be accompanied by release 
of the extra electrostatic energy in the form of kinetic energy of 
ions. Our results allow one to calculate the exact amount of 
energy released in any particular case. For estimates, let us note 
that according to Figs.\ \ref{dU_s} and \ref{dU_c}, about 
$0.0005 Z^2 e^2/a$ per each nucleus can be typically released. Thus 
the total energy release per event can be estimated as
\begin{equation}
        Q \approx 1.4 \times 10^{38} \rho_{11}^{4/3} 
        Z_{44}^2 A^{-4/3}_{126} 
        S_{10} h_3 ~~~ {\rm erg~,}
\end{equation}
where $\rho_{11}$ is the mass density in units of $10^{11}$ g cm$^{-3}$,
$A$ is the ion mass number, $Z_{44}=Z/44$,
$A_{126}=A/126$, $S_{10}$ is the destructed area in units of 1 km$^2$
and $h_{3}$ is the height of the destructed volume in units of 
$10^3$ cm.

In order to produce such a critical stretch at 
$\rho = 10^{11}$ g cm$^{-3}$, $Z=44$, and $A=126$, one has to have 
$B\delta B \sim 10^{28}$ G$^2$ (Paper 1). In other words,
a magnetic field of, say, $2 \times 10^{14}$ G has to vary by 25\% over 
the distance of $\sqrt{S}\sim 1$ km. This seems realistic for 
magnetars.

\section{Conclusions}
\label{Sec_conc}
In an attempt to deepen understanding of the strength of the 
NS crust, we have performed a detailed study of uniaxial stretches 
of crystallites comprising it. The crystallites are  
modelled as arbitrarily oriented perfect bcc Coulomb crystals. 
In Section \ref{Sec_res} we have demonstrated that the critical strain, 
above which the crystal lattice lost its stability, was highly 
anisotropic varying from 0.04 to 0.3 as a function of mutual 
orientation of the stretch direction and the crystallographic axes. 
The same holds true for the critical pressure anisotropy (which varies 
from 0.005 to $0.04 \, nZ^2e^2/a$) and the excess energy per ion at 
the critical strain (which may vary from 0.0001 to 
$0.0018 \, Z^2e^2/a$; note that this is insufficient to melt the whole 
crystallite even if it is already at melting temperature). The high 
degree of breaking anisotropy implies that the von Misses and Tresca 
failure criteria are invalid for bcc Coulomb monocrystals.

Our numerical results are presented in 
Figs.\ \ref{xicrit_s}--\ref{dU_c} in dimensionless form. These data 
can be converted to physical units with the aid of the scaling 
relations (\ref{dP_scale}) and (\ref{U_scale}). All results of 
Section \ref{Sec_res} are based on standard lattice dynamics and are 
rather robust. However, they are obtained for a pure Coulomb static 
lattice. We expect that ion motion and electron screening effects 
would reduce critical deformation parameters by some 10--20 \%. The 
approach of Paper 1 applied here can be used to study the whole 
five-dimensional yield surface in the six-dimensional space of 
strains, while in this paper we constrain ourselves to uniaxial 
deformations only, which span a three-dimensional 
subspace\footnote{As discussed in Section \ref{Sec_res}, a uniform 
compression should not lead to breaking. Hence, two additional 
deformation parameters require further research. These can be, for 
instance, the orientation of the second principal strain axis and the 
strain along this axis.}.

In Section \ref{Sec_spec}, our results for monocrystals are used to 
estimate the strength of the polycrystalline NS crust. We argue that 
polycrystalline crust breaking is determined by either the minimum 
critical strain or the minimum critical pressure anisotropy achievable 
in monocrystals. This implies the breaking strain for the uniaxial 
stretch to be $\epsilon_{\rm crit} \approx$ 0.03--0.04, which is 
a factor of 2.5--3 lower than the 
typically assumed value of $\epsilon^{\rm trad}_{\rm crit} \approx 0.1$ 
motivated by MD simulations of a  
shear deformation. Unlike the breaking strain, the breaking stress is 
found to be model dependent. The pressure anisotropy at breaking 
is predicted to lie in the range between 0.005 and 
$0.014 \, nZ^2e^2/a$. This range excludes all the estimates reported 
previously for the shear deformation: 
$\Delta P^{\rm pMD}_{\rm crit} \approx 0.016$ from 
the polycrystalline MD, $\Delta P^{\rm trad}_{\rm crit} \approx 0.0238$ 
used traditionally, and $\Delta P^{\rm MD}_{\rm crit} \approx 0.039$ 
from the monocrystalline MD. Alternatively, referring to the von 
Misses criterion, the critical parameter $\sigma_{\rm max}$ for 
stretches is predicted to lie in the range 
from 0.003 to 0.008. The upper bound here coincides with 
$\sigma^{\rm pMD}_{\rm max}$ for the shear deformation in the 
polycrystalline MD. The other two estimates for the shear deformation, 
$\sigma^{\rm trad}_{\rm max} \approx 0.0119$ (traditional) and 
$\sigma^{\rm MD}_{\rm max} \approx 0.0195$ (from the monocrystalline 
MD), fall outside our range.    

We also discuss a formation of large crystallites in the NS crust and 
raise a possibility of plastic flow driven by the stress-induced grain 
growth. Finally, we calculate heat release from crust breaking events 
in the context of magnetic field evolution and magnetar flaring 
activity. We would like to emphasise that Section \ref{Sec_spec} is 
of somewhat speculative character. However, its predictions can, 
in principle, be checked with carefully designed MD simulations. 
We plan to do this subsequently.

\section*{Acknowledgments}
We are grateful to Russian Science Foundation grant 14-12-00316 for 
support.

\bibliographystyle{mnras}
\bibliography{literature}

\begin{thebibliography}{}
\makeatletter
\relax
\def\mn@urlcharsother{\let\do\@makeother \do\$\do\&\do\#\do\^\do\_\do\%\do\~}
\def\mn@doi{\begingroup\mn@urlcharsother \@ifnextchar [ {\mn@doi@}
  {\mn@doi@[]}}
\def\mn@doi@[#1]#2{\def\@tempa{#1}\ifx\@tempa\@empty \href
  {http://dx.doi.org/#2} {doi:#2}\else \href {http://dx.doi.org/#2} {#1}\fi
  \endgroup}
\def\mn@eprint#1#2{\mn@eprint@#1:#2::\@nil}
\def\mn@eprint@arXiv#1{\href {http://arxiv.org/abs/#1} {{\tt arXiv:#1}}}
\def\mn@eprint@dblp#1{\href {http://dblp.uni-trier.de/rec/bibtex/#1.xml}
  {dblp:#1}}
\def\mn@eprint@#1:#2:#3:#4\@nil{\def\@tempa {#1}\def\@tempb {#2}\def\@tempc
  {#3}\ifx \@tempc \@empty \let \@tempc \@tempb \let \@tempb \@tempa \fi \ifx
  \@tempb \@empty \def\@tempb {arXiv}\fi \@ifundefined
  {mn@eprint@\@tempb}{\@tempb:\@tempc}{\expandafter \expandafter \csname
  mn@eprint@\@tempb\endcsname \expandafter{\@tempc}}}

\bibitem[\protect\citeauthoryear{{Baiko}}{{Baiko}}{2011}]{Baiko11}
{Baiko} D.~A.,  2011, \mn@doi [\mnras] {10.1111/j.1365-2966.2011.18819.x},
  \href {http://adsabs.harvard.edu/abs/2011MNRAS.416...22B} {416, 22}

\bibitem[\protect\citeauthoryear{{Baiko}}{{Baiko}}{2014}]{Baiko14}
{Baiko} D.~A.,  2014, in Journal of Physics Conference Series. p. 012010,
  \mn@doi{10.1088/1742-6596/496/1/012010}

\bibitem[\protect\citeauthoryear{{Baiko}}{{Baiko}}{2015}]{Baiko15}
{Baiko} D.~A.,  2015, \mn@doi [\mnras] {10.1093/mnras/stv1166}, \href
  {http://adsabs.harvard.edu/abs/2015MNRAS.451.3055B} {451, 3055}

\bibitem[\protect\citeauthoryear{{Baiko} \& {Kozhberov}}{{Baiko} \&
  {Kozhberov}}{2017}]{BK17}
{Baiko} D.~A.,  {Kozhberov} A.~A.,  2017, \mn@doi [\mnras]
  {10.1093/mnras/stx1270}, \href
  {http://adsabs.harvard.edu/abs/2017MNRAS.470..517B} {470, 517}

\bibitem[\protect\citeauthoryear{Caplan \& Horowitz}{Caplan \&
  Horowitz}{2017}]{ch17}
Caplan M.~E.,  Horowitz C.~J.,  2017, \mn@doi [Rev. Mod. Phys.]
  {10.1103/RevModPhys.89.041002}, 89, 041002

\bibitem[\protect\citeauthoryear{{Caplan}, {Cumming}, {Berry}, {Horowitz}  \&
  {Mckinven}}{{Caplan} et~al.}{2018}]{Caplan_etal18}
{Caplan} M.~E.,  {Cumming} A.,  {Berry} D.~K.,  {Horowitz} C.~J.,   {Mckinven}
  R.,  2018, preprint, \href
  {http://adsabs.harvard.edu/abs/2018arXiv180406942C} {} (\mn@eprint {arXiv}
  {1804.06942})

\bibitem[\protect\citeauthoryear{{Chamel} \& {Haensel}}{{Chamel} \&
  {Haensel}}{2008}]{ch08}
{Chamel} N.,  {Haensel} P.,  2008, Liv. Rev. Relativ., \href
  {http://adsabs.harvard.edu/abs/2008LRR....11...10C} {11, 10}

\bibitem[\protect\citeauthoryear{{Chugunov} \& {Horowitz}}{{Chugunov} \&
  {Horowitz}}{2010}]{ch10}
{Chugunov} A.~I.,  {Horowitz} C.~J.,  2010, \mn@doi [\mnras]
  {10.1111/j.1745-3933.2010.00903.x}, \href
  {http://adsabs.harvard.edu/abs/2010MNRAS.407L..54C} {407, L54}

\bibitem[\protect\citeauthoryear{{Chugunov} \& {Horowitz}}{{Chugunov} \&
  {Horowitz}}{2012}]{ch12}
{Chugunov} A.~I.,  {Horowitz} C.~J.,  2012, \mn@doi [Contributions to Plasma
  Physics] {10.1002/ctpp.201100075}, \href
  {http://adsabs.harvard.edu/abs/2012CoPP...52..122C} {52, 122}

%
\bibitem[\protect\citeauthoryear{{Eshelby}}{{Eshelby}}{1961}]{Eshelby61}
{Eshelby} J.~D.,  1961, Prog.\ Solid Mech., 2, 87

\bibitem[\protect\citeauthoryear{{Gabler}, {Cerd{\'a}-Dur{\'a}n},
  {Stergioulas}, {Font}  \& {M{\"u}ller}}{{Gabler}
  et~al.}{2018}]{Gabler_etal18}
{Gabler} M.,  {Cerd{\'a}-Dur{\'a}n} P.,  {Stergioulas} N.,  {Font} J.~A.,
  {M{\"u}ller} E.,  2018, \mn@doi [\mnras] {10.1093/mnras/sty445}, \href
  {http://adsabs.harvard.edu/abs/2018MNRAS.476.4199G} {476, 4199}

\bibitem[\protect\citeauthoryear{Haensel, Potekhin  \& Yakovlev}{Haensel
  et~al.}{2007}]{hpy07}
Haensel P.,  Potekhin A.,   Yakovlev D.,  2007, Neutron Stars 1: Equation of
  State and Structure.
Astrophysics and Space Science Library, Springer

\bibitem[\protect\citeauthoryear{{Hansen}}{{Hansen}}{2004}]{H04}
{Hansen} B.,  2004, \mn@doi [\physrep] {10.1016/j.physrep.2004.07.001}, \href
  {http://adsabs.harvard.edu/abs/2004PhR...399....1H} {399, 1}

\bibitem[\protect\citeauthoryear{{Haskell} \& {Patruno}}{{Haskell} \&
  {Patruno}}{2017}]{hp17}
{Haskell} B.,  {Patruno} A.,  2017, \mn@doi [Physical Review Letters]
  {10.1103/PhysRevLett.119.161103}, \href
  {http://adsabs.harvard.edu/abs/2017PhRvL.119p1103H} {119, 161103}

\bibitem[\protect\citeauthoryear{{Hoffman} \& {Heyl}}{{Hoffman} \&
  {Heyl}}{2012}]{hh12}
{Hoffman} K.,  {Heyl} J.,  2012, \mn@doi [\mnras]
  {10.1111/j.1365-2966.2012.21921.x}, \href
  {http://adsabs.harvard.edu/abs/2012MNRAS.426.2404H} {426, 2404}

\bibitem[\protect\citeauthoryear{{Horowitz}}{{Horowitz}}{2010}]{Horowitz10}
{Horowitz} C.~J.,  2010, \mn@doi [\prd] {10.1103/PhysRevD.81.103001}, \href
  {http://adsabs.harvard.edu/abs/2010PhRvD..81j3001H} {81, 103001}

\bibitem[\protect\citeauthoryear{{Horowitz} \& {Kadau}}{{Horowitz} \&
  {Kadau}}{2009}]{hk09}
{Horowitz} C.~J.,  {Kadau} K.,  2009, \mn@doi [Physical Review Letters]
  {10.1103/PhysRevLett.102.191102}, \href
  {http://adsabs.harvard.edu/abs/2009PhRvL.102s1102H} {102, 191102}

\bibitem[\protect\citeauthoryear{{Johnson-McDaniel}}{{Johnson-McDaniel}}{2013}]{jm13}
{Johnson-McDaniel} N.~K.,  2013, \mn@doi [\prd] {10.1103/PhysRevD.88.044016},
  \href {http://adsabs.harvard.edu/abs/2013PhRvD..88d4016J} {88, 044016}

\bibitem[\protect\citeauthoryear{{Johnson-McDaniel} \&
  {Owen}}{{Johnson-McDaniel} \& {Owen}}{2013}]{jmo13}
{Johnson-McDaniel} N.~K.,  {Owen} B.~J.,  2013, \mn@doi [\prd]
  {10.1103/PhysRevD.88.044004}, \href
  {http://adsabs.harvard.edu/abs/2013PhRvD..88d4004J} {88, 044004}

\bibitem[\protect\citeauthoryear{{Kobyakov} \& {Pethick}}{{Kobyakov} \&
  {Pethick}}{2015}]{kp15}
{Kobyakov} D.,  {Pethick} C.~J.,  2015, \mn@doi [\mnras]
  {10.1093/mnrasl/slv027}, \href
  {http://adsabs.harvard.edu/abs/2015MNRAS.449L.110K} {449, L110}

\bibitem[\protect\citeauthoryear{{Kocks}}{{Kocks}}{1970}]{Kocks70}
{Kocks} U.~F.,  1970, \mn@doi [Metallurgical Transactions]
  {10.1007/BF02900224}, \href
  {http://adsabs.harvard.edu/abs/1970MT......1.1121K} {1, 1121}

\bibitem[\protect\citeauthoryear{Landau, Lifshitz, Kosevich, Sykes,
  Pitaevski{\u\i}  \& Reid}{Landau et~al.}{1986}]{ll_elast}
Landau L.,  Lifshitz E.,  Kosevich A.,  Sykes J.,  Pitaevski{\u\i} L.,   Reid
  W.,  1986, Theory of Elasticity.
Course of theoretical physics, Butterworth-Heinemann, \url
  {https://books.google.ru/books?id=tpY-VkwCkAIC}

\bibitem[\protect\citeauthoryear{{Lander}}{{Lander}}{2016}]{Lander16}
{Lander} S.~K.,  2016, \mn@doi [\apjl] {10.3847/2041-8205/824/2/L21}, \href
  {http://adsabs.harvard.edu/abs/2016ApJ...824L..21L} {824, L21}

\bibitem[\protect\citeauthoryear{{Li}, {Levin}  \& {Beloborodov}}{{Li}
  et~al.}{2016}]{llb16}
{Li} X.,  {Levin} Y.,   {Beloborodov} A.~M.,  2016, \mn@doi [\apj]
  {10.3847/1538-4357/833/2/189}, \href
  {http://adsabs.harvard.edu/abs/2016ApJ...833..189L} {833, 189}

\bibitem[\protect\citeauthoryear{Masuda, Kanazawa, Tobe  \& Sato}{Masuda
  et~al.}{2018}]{GrainRotationAndSliding}
Masuda H.,  Kanazawa T.,  Tobe H.,   Sato E.,  2018, \mn@doi [Scripta
  Materialia] {https://doi.org/10.1016/j.scriptamat.2018.02.021}, 149, 84

\bibitem[\protect\citeauthoryear{{Ogata} \& {Ichimaru}}{{Ogata} \&
  {Ichimaru}}{1990}]{oi90}
{Ogata} S.,  {Ichimaru} S.,  1990, \mn@doi [\pra] {10.1103/PhysRevA.42.4867},
  \href {http://adsabs.harvard.edu/abs/1990PhRvA..42.4867O} {42, 4867}

\bibitem[\protect\citeauthoryear{{Regel'}, {Slutsker}  \& {Tomashevski{\u
  i}}}{{Regel'} et~al.}{1972}]{rst72}
{Regel'} V.~R.,  {Slutsker} A.~I.,   {Tomashevski{\u i}} {\'E}.~E.,  1972,
  \mn@doi [Soviet Physics Uspekhi] {10.1070/PU1972v015n01ABEH004945}, \href
  {http://adsabs.harvard.edu/abs/1972SvPhU..15...45R} {15, 45}

\bibitem[\protect\citeauthoryear{Rupert, Gianola, Gan  \& Hemker}{Rupert
  et~al.}{2009}]{grain_grow_Science2009}
Rupert T.~J.,  Gianola D.~S.,  Gan Y.,   Hemker K.~J.,  2009, \mn@doi [Science]
  {10.1126/science.1178226}, 326, 1686

\bibitem[\protect\citeauthoryear{{Strohmayer}, {Ogata}, {Iyetomi}, {Ichimaru}
  \& {van Horn}}{{Strohmayer} et~al.}{1991}]{Strohmayer_etal91}
{Strohmayer} T.,  {Ogata} S.,  {Iyetomi} H.,  {Ichimaru} S.,   {van Horn}
  H.~M.,  1991, \mn@doi [\apj] {10.1086/170231}, \href
  {http://adsabs.harvard.edu/abs/1991ApJ...375..679S} {375, 679}

\bibitem[\protect\citeauthoryear{{Tsang}, {Read}, {Hinderer}, {Piro}  \&
  {Bondarescu}}{{Tsang} et~al.}{2012}]{Tsang_etal12}
{Tsang} D.,  {Read} J.~S.,  {Hinderer} T.,  {Piro} A.~L.,   {Bondarescu} R.,
  2012, \mn@doi [Physical Review Letters] {10.1103/PhysRevLett.108.011102},
  \href {http://adsabs.harvard.edu/abs/2012PhRvL.108a1102T} {108, 011102}

\bibitem[\protect\citeauthoryear{Ushomirsky, Cutler  \& Bildsten}{Ushomirsky
  et~al.}{2000}]{ucb00}
Ushomirsky G.,  Cutler C.,   Bildsten L.,  2000, \mn@doi [Monthly Notices of
  the Royal Astronomical Society] {10.1046/j.1365-8711.2000.03938.x}, 319, 902

\makeatother
\end{thebibliography}

\label{lastpage}
\end{document}